\documentclass{emulateapj}
\usepackage{graphics,graphicx}
\usepackage{todonotes}
\def\kms{km\,s$^{-1}$}
\def\etal{ et~al.\,}

\def\spitzer{{\it Spitzer}}

\def\chan{{\it Chandra}}
\def\wifes{WiFeS\,}

\tighten
\begin{document}
\submitted{Accepted for publication in ApJ}
\title{Integral Field Spectroscopy of Balmer-Dominated Shocks in the Magellanic Cloud Supernova Remnant N103B}

\author{Parviz Ghavamian\altaffilmark{1},  Ivo R.~Seitenzahl\altaffilmark{2,3,4}, Fr{\'e}d{\'e}ric P.~A. Vogt\altaffilmark{5 $\dagger$}, M. A. Dopita\altaffilmark{2}, Jason P. Terry\altaffilmark{6}, Brian J. Williams\altaffilmark{7,9}, P. Frank Winkler\altaffilmark{8} 
  }

\altaffiltext{1}{Department of Physics, Astronomy and Geosciences,
  Towson University, Towson, MD, 21252; pghavamian@towson.edu}
\altaffiltext{2}{Research School of Astronomy and Astrophysics, Australian National University, Canberra, ACT 2611, Australia}
\altaffiltext{3}{ARC Centre for All-sky Astrophysics (CAASTRO).}
\altaffiltext{4}{School of Physical, Environmental and Mathematical Sciences, University of New South Wales, Australian Defence Force Academy, Canberra, ACT 2600, Australia}
\altaffiltext{5}{European Southern Observatory, Av. Alonso de C\'ordova 3107, 763 0355 Vitacura, Santiago, Chile. }
\altaffiltext{$\dagger$}{ESO Fellow.}
\altaffiltext{6}{Department of Physics and Astronomy, University of Georgia, USA.}
\altaffiltext{7}{NASA Goddard Space Flight Center, Greenbelt, MD, 20771}
\altaffiltext{8}{Department of Physics, Middlebury College, Middlebury, VT, 05753}
\altaffiltext{9}{Space Telescope Science Institute, 3700 San Martin Drive, Baltimore, MD, 21218}

\begin{abstract}
We present results of integral field spectroscopy of Balmer-dominated shocks in the LMC supernova remnant (SNR) N103B, carried out using the Wide Field Integral Spectrograph (\wifes) on the 2.3\,m telescope at Siding Spring Observatory in Australia.  Existing X-ray studies of N103B have indicated a Type Ia supernova origin.  Radiative shock emission from clumpy material surrounding the SNR may result from interaction of the forward shock with relic stellar wind material, possibly implicating a thermonuclear explosion in a single-degenerate binary system.  The recently discovered Balmer-dominated shocks mark the impact of the forward shock with low density, partially neutral CSM gas, and form a partial shell encircling clumps of material exhibiting radiative shocks.   The \wifes spectra of N103B reveal broad H$\alpha$ emission having a width as high as 2350 \kms\, along the northern rim, and both H$\alpha$ and H$\beta$ broad profiles having width around 1300 \kms\, along the southern rim.  Fits to the H$\alpha$ line profiles indicate that in addition to  the usual broad and narrow emission components, a third component of intermediate width exists in these Balmer-dominated shocks, ranging from around 125 \kms\, up to 225 \kms\, in width.  This is consistent with predictions of recent Balmer-dominated shock models, which predict that an intermediate-width component will be generated in a fast neutral precursor.  We derive a Sedov age of approximately 685$\pm$20 years for N103B from the Balmer-dominated spectra, consistent with the young age of 380 -- 860 years estimated from light echo studies.   

\end{abstract}


\section{Introduction}

Type Ia supernovae are now widely understood to arise from the thermonuclear destruction of a white dwarf star, but the exact trigger mechanism for the explosion is still unknown.  Type Ia supernovae are believed to be caused either by coalescence of two white dwarf stars (the so-called double-degenerate scenario; Iben \& Tutukov 1984; Webbink 1984), or by an alternative mechanism, runaway nuclear burning caused by accretion from a main sequence or giant star onto a white dwarf (single-degenerate scenario, Whelan \& Iben 1973). Conclusive evidence in favor of the single-degenerate mechanism has been difficult to find.  One of the most promising methods used so far is to search for evidence of interaction between the supernova remnant and a dense circumstellar wind, of the type generated prior to the explosion by the donor star.  One promising candidate has been Kepler's supernova remnant, wherein the Balmer-dominated shocks are interacting with dense clumps of material exhibiting radiative shocks  (Blair \etal\, 1991; Sankrit \etal\, 2016).  There is evidence of a silicate-rich dust composition in the shocked material, possibly indicating an origin in either the stellar wind of the non-degenerate donor star or the progenitor itself (Williams\etal  2012).   

Recently, a second  SNR has been identified as a candidate for a single-degenerate Type Ia:  N103B (SNR 0509$-$68.7) in the Large Magellanic Cloud (Williams\etal  2014). X-ray analyses of N103B (Hughes \etal 1995; Lewis\etal 2003; Lopez\etal 2011; Yamaguchi\etal 2014) indicate an Fe/O abundance and morphology consistent with a Type Ia explosion.  In optical images the SNR exhibits a prominent collection of compact, radiatively shocked ejecta knots emitting in H$\alpha$, [S\,\textsc{ii}] and [O\,\textsc{iii}] (Williams\etal  2014; Li \etal\, 2017) embedded within a partially formed shell of H$\alpha$ emission.   \spitzer\, IRS spectroscopy of N103B (Williams\etal  2014) shows a strong silicate bump close to 18 $\micron$ and a lack of nitrogen enhancement in the knots, suggesting the presence of an unevolved donor star.
The small angular size of this SNR ($\sim$30$^{\prime\prime}$ across, or 7.2 pc at the LMC distance of 50 kpc) indicates a SNR similar in size to Tycho's SNR.   However, light echoes from N103B place its age at approximately 860 years (Rest et al. 2005), or about twice as old as Tycho and Kepler.  Recently, Li\etal  (2017) have presented both narrowband and broadband imagery of N103B taken with the WFC3 instrument on HST.  They have identified a potential main sequence surviving companion near the center of the SNR, providing further evidence for a single-degenerate explosion scenario.  The HST images showed that  most of the filaments surrounding N103B lack forbidden line emission, and exhibit the delicate morphology characteristic of Balmer-dominated shocks.   Ground based optical spectra presented by Li \etal\, (2017) revealed highly redshifted broad H$\alpha$ emission ($\sim$500 \kms) from knots near the projected center of N103B, further evidence of Balmer-dominated shocks in this SNR.    
  
In this paper we present integral field spectra of Balmer-dominated shocks in N103B acquired with the Wide Field Imaging Spectrograph (\wifes; Dopita\etal  2007, 2010) on the 2.3\,m telescope of the Australian National University.  Our observations cover the spectral range 3500 -- 7000 \AA\, and have lower spectral resolution than the echelle spectra of Li \etal\,(2017), allowing for detection of the full extent of the broad H$\alpha$ line both in the interior and along the southern and northern edges of N103B.  In at least one location the H$\beta$ broad component is also detected, making N103B only the fifth SNR (after Tycho, SN 1006, RCW 86 and the Cygnus Loop; Ghavamian\etal  2001; 2002) where broad components are detected in both Balmer lines. We also detect regions transitioning from Balmer-dominated shocks to radiative shocks, where both broad H$\alpha$ and moderate forbidden line emission is present.  

First described by Chevalier, Kirshner \& Raymond (1980), each Balmer line consists of a narrow component produced when cold ambient neutrals are overrun by the shock, and a broad component produced when charge exchange with fast protons generates hot neutral hydrogen behind the shock (see reviews by Heng 2010 and by Ghavamian\etal  2013, and references in these papers). The width of the broad component is determined by the postshock proton temperature and the cosmic ray acceleration efficiency, while the ratio of broad-to-narrow flux is sensitive to those parameters as well as the electron-proton temperature equilibration and, to a lesser extent, the preshock neutral fraction (Morlino\etal 2012, 2013).  Analysis of the Balmer line profiles gives valuable information on the kinematics of the supernova remnant, as well as insight into the nature of collisionless heating processes operating in high Mach number nonradiative shocks. (Smith \etal\, 1991, Heng \etal\, 2010, Ghavamian \etal\, 2001, 2002, 2003). 

One advantage of the \wifes\ integral field observations, aside from full spatial coverage, is its combination of high throughput and high spectral resolution (FWHM 43 \kms\ at H$\alpha$, and 100 \kms\ at H$\beta$). The simultaneous recording of the blue and red spectra on two separate CCDs allows for moderately high dispersion spectra to be acquired over a broad wavelength range. One result of our observations with this instrument is the detection of the intermediate velocity H$\alpha$ component in N103B as predicted by the latest models of Balmer-dominated shocks (Morlino\etal  2012, 2013).  This component has a velocity intermediate between that of the narrow component (tens of \kms) and the broad component (thousands of \kms), and is produced by the heating of the preshock gas when broad component neutrals escape upstream and undergo charge exchange with cold protons (i.e., the `neutral return flux', Morlino\etal  2013). Our H$\alpha$ line profile fits for N103B show intermediate component widths $\sim$125 -- 225 \kms, consistent with the range predicted by theoretical models of Morlino\etal  (2013).  However, the ratio of the broad-to-narrow components falls well below the latest theoretically predicted ratios of Morlino\etal  (2012).  We describe these results, as well as possible suggestions for future modeling efforts, in this paper.

\section{Observations and Data Reduction}

 The \wifes observations of N103B were performed on the nights of 2014 December 18 and 19 (PI Seitenzahl). \wifes is located at the Nasmyth A focus of the Australian National University 2.3m telescope located at Siding Spring Observatory, Australia.  The spectra were acquired in `binned mode,' providing a field-of-view of $25 \times 38$ spatial pixels (spaxels) each 1$^{\arcsec}\times$1$^{\arcsec}$ in angular size. The instrument is a double-beam spectrograph providing independent channels for each of the blue and the red wavelength ranges. We used the B3000 and R7000 gratings, covering together (and simultaneously) the 3500 -- 7000 \AA\, wavelength range, and giving a resolution of R = 3000 in the blue (3500 -- 5700 \AA) and R = 7000 in the red (5300 -- 7000 \AA).  The corresponding velocity resolutions are 100 \kms\ in H$\beta$, and 43 \kms\ in H$\alpha$. 

Since N103B slightly exceeds the \wifes\ field-of-view, we observed the SNR in a mosaic of two overlapping fields. For each field we acquired two 1800s exposures with the center of N103B falling into the overlap region of the two fields. Two 900\,s sky reference frames were acquired for each field, which were scaled and subtracted from the two co-added frames for each field during subsequent data reduction.  We separately corrected for the OH and H$_2$O telluric absorption features in each frame, then used the STIS spectrophotometric standard HD261696 to perform the absolute photometric calibration of the data cubes. Transparency conditions varied by less than 10\% during the N103B observations and the seeing varied between 1.5$^{\arcsec}$ and 2.5$^{\arcsec}$.  We regularly acquired internal wavelength calibration and bias frames during the observations, and obtained internal continuum lamp flat fields and twilight sky flats in order to provide sensitivity corrections in both the spectral and spatial directions.

The data were reduced using \textsc{pywifes} v0.7.3 reduction pipeline (Childress et al. 2014). The pipeline produced a wavelength calibrated, sensitivity corrected, photometrically calibrated data cube corrected for telluric features and cosmic ray events. The \textsc{pywifes} pipeline uses an optical model of the spectrograph to provide the wavelength calibration, resulting in a wavelength solution valid across the entire detectors.  In doing so, each pixel on the CCD is assigned a precise wavelength and spatial coordinate (i.e., a spaxel) by the data reduction pipeline.  The final data cubes (one per exposure and per spectrograph arm) are then reconstructed and interpolated onto a regular three-dimensional grid with wavelength intervals of 0.768 \AA\, in the blue and 0.439 \AA\, in the red. 

\begin{figure}[htb!] 
\centering{\includegraphics[scale=0.7]{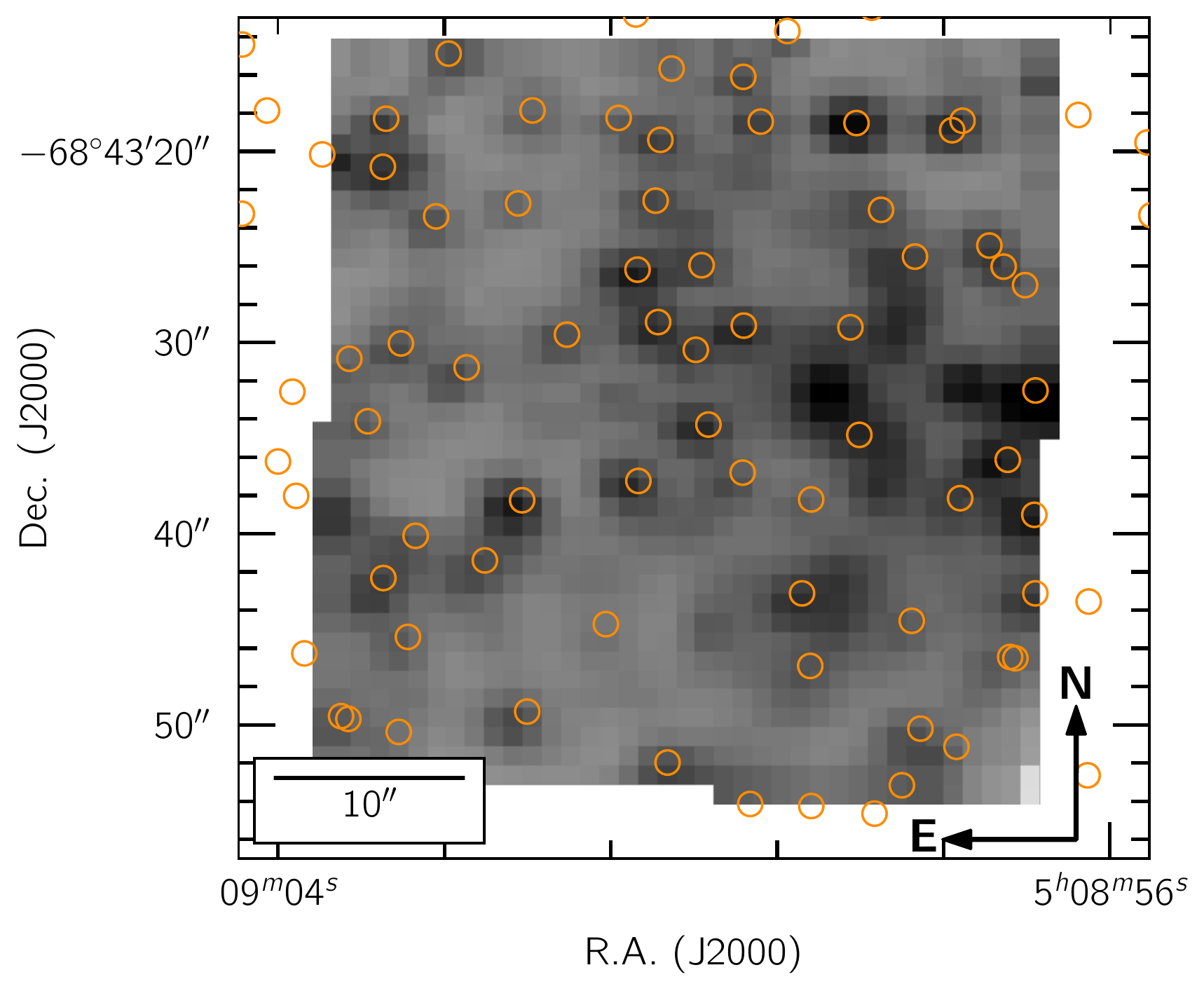}}
  \caption{  Comparison between the reconstructed R-band image from the WiFeS mosaic of N103B (greyscale) and the locations of stars from the GAIA survey (orange circles).  The absolute WCS accuracy of the WiFeS mosaic with respect to GAIA, and the respective positioning accuracy between the WiFeS fields, is $\lesssim1$\arcsec.  }
   \label{fig:fig1}
\end{figure}

The red and blue final mosaics, combining 4 individual reduced cubes each, were assembled using a custom \textsc{python} script\footnote{The script was written by F.P.A. Vogt and is available on request.}. The respective alignment of the different \wifes fields in the mosaic was derived by comparing the reconstructed summed continuum frames from the red cubes with the Digitized Sky Survey 2 (DSS 2) red band image of the area. Given the mean seeing conditions during the observations ($\sim$2.0$^{\arcsec}$), the spatial shifts between each individual cube are rounded to the nearest integer for simplicity, and also to avoid superfluous resampling of the data. All data cubes are forced onto the same wavelength grid in the data reduction process, so that no shift is required along that axis. The final mosaic has dimensions 40$\arcsec\,\times\,$36$\arcsec$ on the sky, which for an assumed distance of the LMC of 50 kpc corresponds to a field of 9.7$\times$8.7 pc. The WCS information in the FITS headers was pegged to the coordinates of a star from the 2MASS catalogue present in the mosaic field (see Figure~\ref{fig:fig1}). The overall positioning of the mosaic and the respective alignment of each field is accurate to  $\lesssim$1\arcsec. 

A final processing step was to remove the contribution of continuum emission from stars in the field. A piece-wise linear function was fit to the stellar continuum in each spaxel and then subtracted from the spectrum in order to bring the background to zero.  This procedure was applied to both blue and red channel data, independently. 

\section{Detection of Broad Balmer Line Emission}

\begin{figure}[htb!] 
\centering{\includegraphics[scale=0.7]{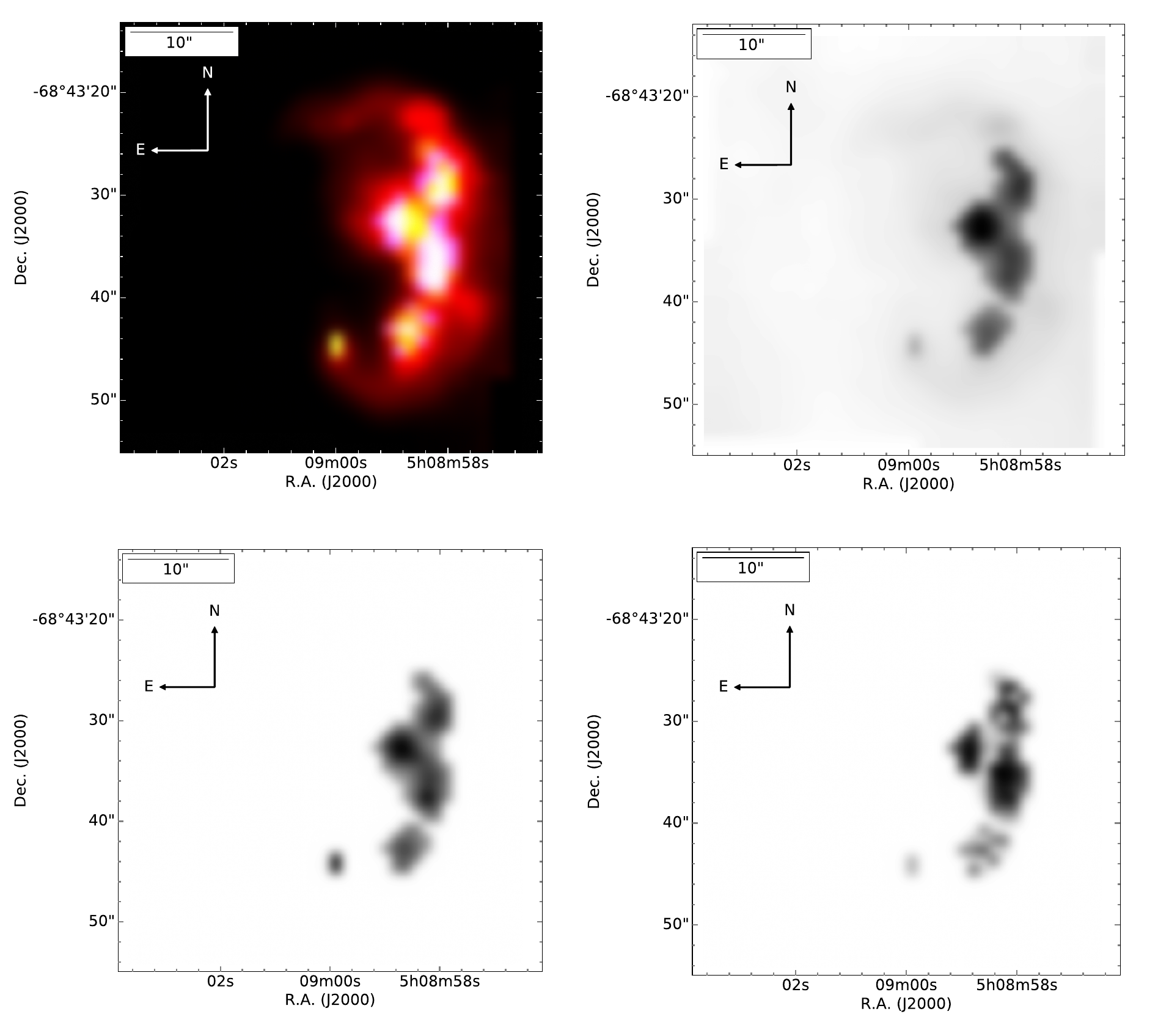}}
   \caption{ Upper left: RGB image of N103B formed from the red and blue channel \wifes mosaics in the emission lines of H$\alpha$ (red), [O\,\textsc{i}] $\lambda$6300 (green) and [Fe\,\textsc{xiv}] $\lambda$5303 (blue).  The [O\,\textsc{i}] image traces the locations of complete radiative shocks, while [Fe\,\textsc{xiv}] is brightest at locations where very strong shocks ($v_{sh}\,\gtrsim\,$300 \kms) are beginning to transition from the non-radiative to radiative phase.  The pure red emission arises from Balmer-dominated shocks.  The ranges from lowest to highest surface brightness are, in units of 10$^{-16}$ ergs cm$^{-2}$ s$^{-1}$ arcsec$^{-2}$ : Red (H$\alpha$):  (0.77 - 4.9), Green [O\,\textsc{i}] (0.03 - 2.2), Blue [Fe\,\textsc{xiv}]: (0.03 to 0.292).  Upper right: H$\alpha$ emission component.  Lower left: [O\,\textsc{i}] $\lambda$6300 emission component.  Lower right: [Fe\,\textsc{xiv}] $\lambda$5303 emission component.  }
   \label{fig:fig2}
\end{figure}

At visible wavelengths N103B appears as a partially complete shell with a shape reminiscent of the numeral `3' (Figure~\ref{fig:fig2}). It is located nearly 4.5$\arcmin$ northeast of the double star cluster NGC 1850, with the closed side of the shell facing the cluster.  The closed side of the shell exhibits faint H$\alpha$ emission, with several bright radiatively shocked clumps in the projected interior. In Figure~\ref{fig:fig2}, we show a 3-color RGB image of N103B in the emission lines of H$\alpha$ (red), [O\,\textsc{i}] $\lambda$6300 (green) and [Fe\,\textsc{xiv}] $\lambda$5303 (blue).  The image, extracted from the \wifes mosaic using custom \textsc{python} software, shows the intensities of the three lines calculated by using the routine MPFIT (Markwardt 2009) to fit multicomponent Gaussian lines in each spaxel (as such, the image is not equivalent to a narrow band image). The RGB map shows extensive Balmer-dominated shock emission along the edges of the `3' pattern, as evidenced by the lack of strong forbidden line emission over much of the shell (Williams \etal\, 2014; Li \etal\, 2017).

However, the shell emission includes a minor contribution from [O\,\textsc{i}] $\lambda$6300 \AA, at a level above the sky background (the orange/yellow emission).  The mixture of [O\,\textsc{i}] $\lambda$6300 \AA\, alongside Balmer-dominated shocks is not easily explained; as we show in the next section, the Balmer-dominated shocks exhibit
broad components well above 1000 \kms, virtually excluding the likelihood that the faint [O\,\textsc{i}] arises from the far cooling zones of radiative shocks. On the other hand, the [O\,\textsc{i}] line width significantly exceeds the instrumental resolution, suggesting that it is nonetheless associated with shocks.
One possibility is that it is excited in the cosmic ray precursors of the Balmer-dominated shocks.  The electron temperatures in the cosmic ray precursors are still not well constrained (Morlino \etal\, 2012) due to uncertainty in the efficiency of wave dissipation driven by the cosmic rays.  Electron temperatures of a few $\times 10^4$ K  are quite plausible ahead of the shock and may cause the [O~I] excitation seen in the WiFeS spectra.  Interestingly, Balmer-dominated shocks with co-extensive [O\,\textsc{i}] have been detected before in circumstellar knots of the other most promising candidate for a single-degenerate Type Ia SN: Kepler's supernova remnant (Knots D41-D45; Blair\etal  1991).  Although an interesting subject, deeper investigation into the nature of the extensive [O\,\textsc{i}] emission is beyond the scope of this paper.  

The coronal [Fe\,\textsc{xiv}] $\lambda$5303 \AA\, emission (bluish regions in Figure~\ref{fig:fig2}) traces radiative and partially radiative shocks with speeds in the 350 -- 450 \kms\, range (Dopita\etal  2016).   Emission from this line surrounds the brightest radiatively shocked clump (which appears in white) near the center of N103B.   The morphology of the emission surrounding the bright clump may reflect density variations in material interacting with the forward shock of N103B.  Invoking ram pressure conservation ($\rho\,V_{sh}^2$\,=\,constant), the denser material at the center of the clump forms fully radiative shocks due to the slower shock speeds and shorter cooling timescales there, while the less dense material surrounding the clump forms strong, partially radiative shocks with long cooling timescales that is responsible for the strong [Fe\,\textsc{xiv}] emission.  

\begin{figure}[ht] 
 \centering

\includegraphics[width=6in]{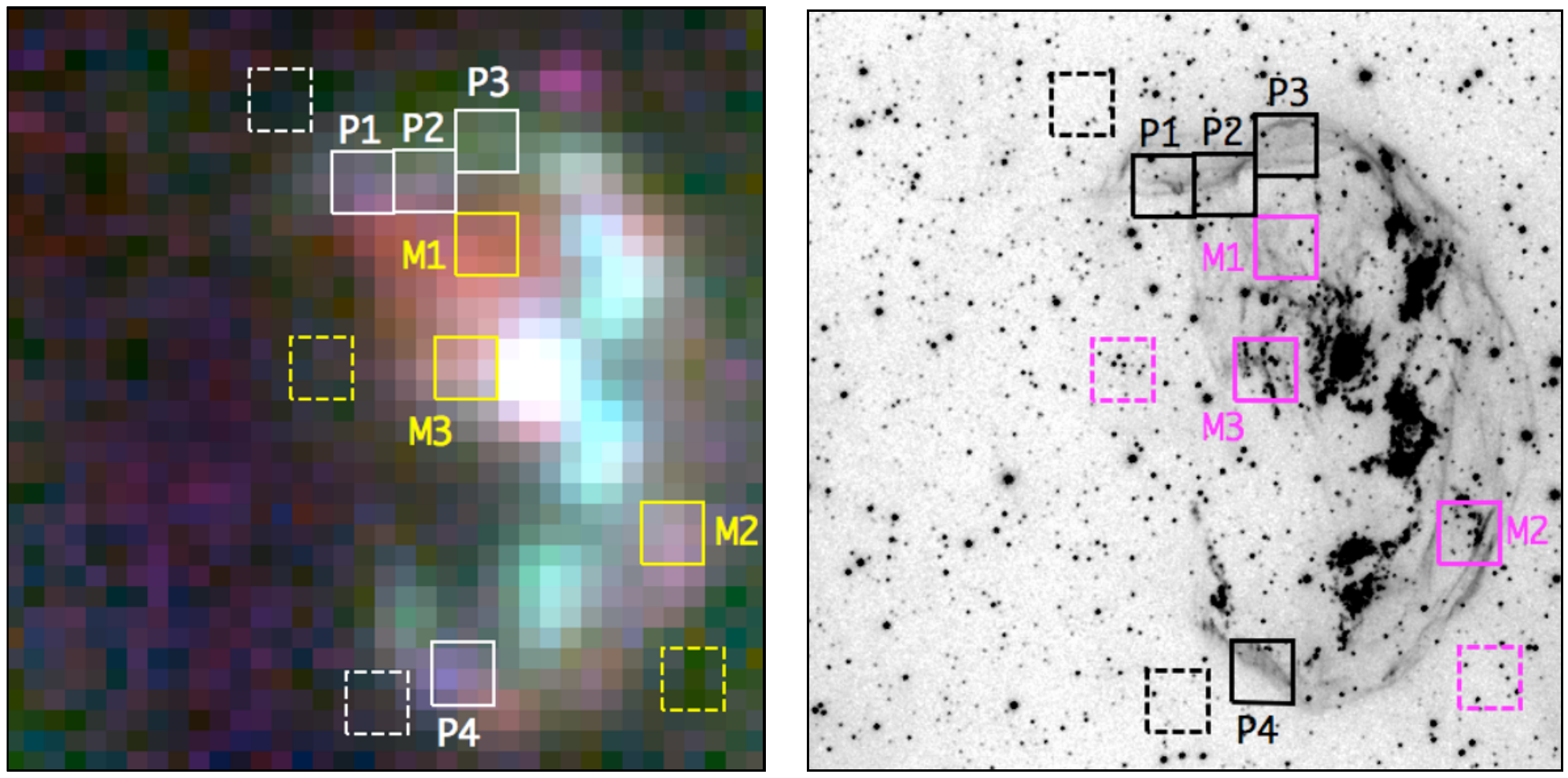}  
      \caption{{\it Left:} Continuum-subtracted line map of N103B, obtained by summing the data cube over three different spectral regions around the H$\alpha$ line at the LMC rest velocity (6569.8 \AA).   
      Red indicates redshifted emission summed between +285 \kms\, and +700 \kms, green indicates emission summed $\pm$185 \kms\, around the center of narrow H$\alpha$, and blue indicates emission summed between $-$285 \kms\, and $-$700 \kms.  Particularly prominent is a patch of redshifted emission in the interior, marking Balmer-dominated shocks on the far side of the SNR.   {\it Right:} HST WFC3 image of N103B in H$\alpha$ (Li \etal\, 2017) , showing the intricate and clumpy distribution of radiative and Balmer-dominated shocks.  
Locations of \wifes spectral extraction boxes of Balmer-dominated shocks (solid boxes P1-P4) and nearby off-source regions used for background subtraction of the spectra (dashed boxes).  Regions are 3$\times$3 pixels (3$\arcsec\,\times$\,3$\arcsec$) in size.  The dashed box near P1, P2 and P3 is used for sky subtraction of the pure Balmer-dominated spectra along the northern edge of N103B (Figures \ref{fig:fig5} and \ref{fig:fig7} ), while the dashed box near P4 is used for sky subtraction along the southern edge.  The dashed box near M1 and M2 is used for sky subtraction of the radiative/Balmer-dominated mixed  spectra, while the dashed box near M3 is used for sky subtraction of that aperture (Figure~\ref{fig:fig4}).  Note that box M1 most closely aligns with the redshifted broad component detected by Li  \etal\, (2017).   }
   \label{fig:fig3}
\end{figure}

A significant portion of the H$\alpha$ emission in N103B is redshifted relative to the LMC rest velocity.  Figure~\ref{fig:fig3} (left panel) shows an RGB  image of N103B obtained by summing H$\alpha$ emission of the \wifes\, data cube between three spectral regions (i.e., line maps).  In red we show emission between +285 \kms\ and +700 \kms\ relative to the local rest wavelength of LMC H$\alpha$ at 6569.8 \AA.  (This velocity interval was chosen to avoid emission from the blue shoulder of the [N\,\textsc{ii}] $\lambda$6583 line, while showing enough redshifted H$\alpha$ to reveal the fainter broad component emission from the Balmer-dominated shocks.)  In green is H$\alpha$ emission between $-185$ \kms\ and $+185$ \kms, and in blue is  blueshifted H$\alpha$, between $-$285 \kms\, and $-$700 \kms.   Extraction boxes for one-dimensional spectra discussed in this paper are marked on the RGB image.  The same apertures are shown on an H$\alpha$ image of N103B acquired with the WFC3 imager on the Hubble Space Telescope (PID 14359; Li \etal\, 2017).  In extracting the spectra we averaged the counts in each frame (wavelength) of the cube within the square shaped regions, 3$\times$3 pixels (3$\arcsec\,\times\,$3$\arcsec$) in size. For each aperture we then obtained a corresponding averaged spectrum of nearby background (marked with dashed boxes in Figure~\ref{fig:fig3}), which we scaled and subtracted from the object spectrum.   

\begin{figure}[ht] 
 \centering
\includegraphics[width=7in]{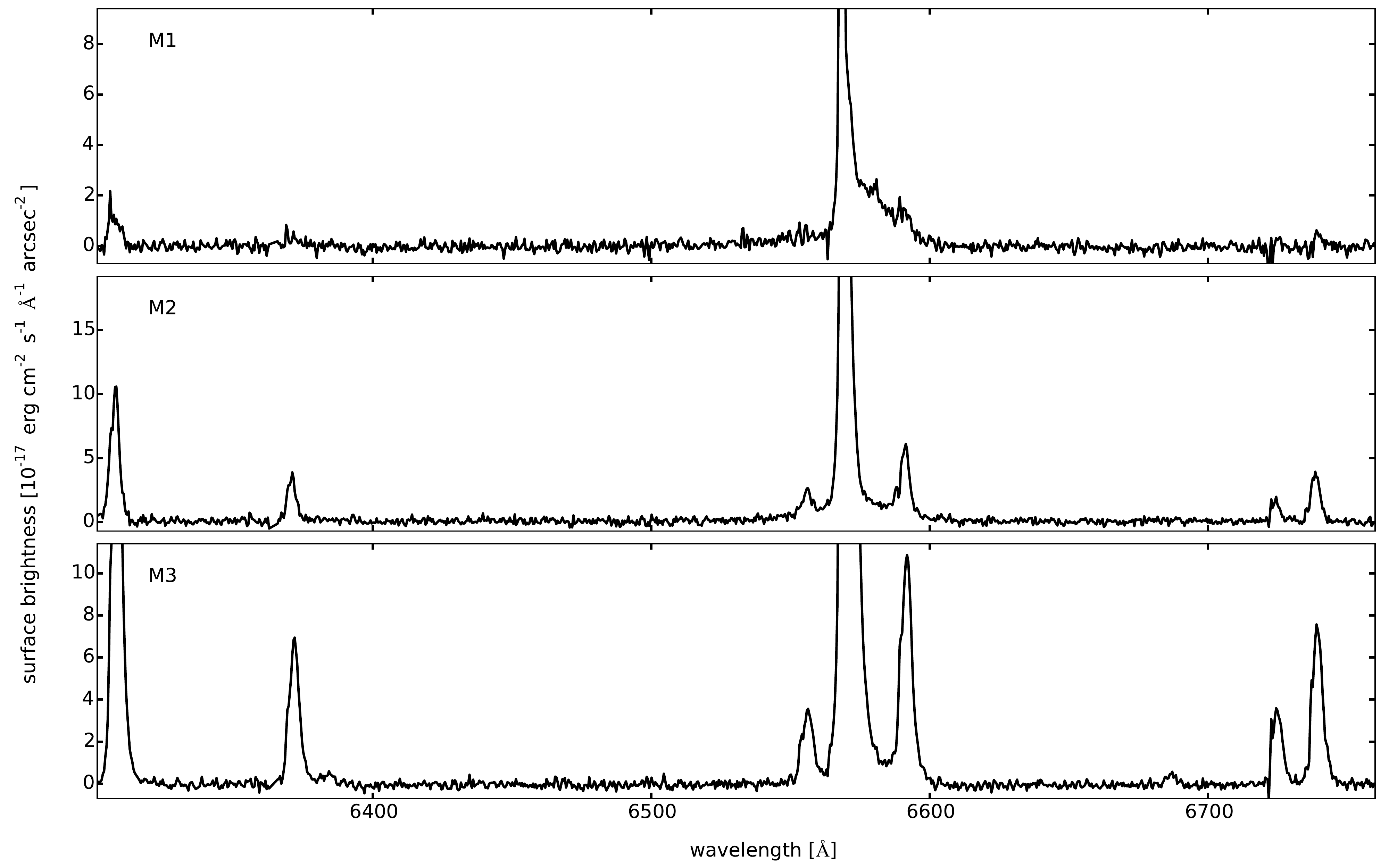}  
   \caption{ Background-subtracted spectra from three positions (marked on Figure~\ref{fig:fig3}) marking varying degrees of radiative shock contribution in N103B.  Top: a Balmer-dominated spectrum with forbidden lines of [O\,\textsc{i}], [N\,\textsc{ii}] and [S\,\textsc{ii}].  The broad H$\alpha$ width is $\sim$2000 \kms, indicating a shock speed too high for radiative cooling, so the forbidden lines may arise from atmospheric smearing of nearby radiative shock emission and/or superposition of multiple shocks along the line of sight. Middle: Another spectrum showing a mix of broad H$\alpha$ and forbidden line emission, this time with multiple velocity components.   Bottom:  An even more extreme case, with broad H$\alpha$ nearly absent absent and with brighter forbidden line components.  }
   \label{fig:fig4}
\end{figure}

Our rationale for the specific sizes, locations and limited number of extraction apertures is as follows:   The spectra of shocks in the N103B interior show considerable spatial variability, making a straightforward characterization of their properties complicated. The type of spectra detected near and within the clumps include forbidden line emission of width $\sim$300 \kms\, mixed with varying amounts of broad Balmer line emission exceeding 2000 \kms\, in width (Figure~\ref{fig:fig4}).  This indicates that both radiative and non-radiative shock emission is detected at these locations. It is uncertain whether the different emission components are merely the result of superposition of multiple types of shocks along the line of sight, or whether they are the result of a range of shock types within the same set of clumps.  However, it is clear that the coarse pixel scale of \wifes\, (1$\arcsec$) and  the smearing effect of atmospheric seeing ($\sim$1.5 -- 2$\arcsec$), make the interpretation of these spectra even more uncertain.   We chose the
3-pixel extraction boxes as a compromise between the coarse pixel size and seeing, the small spatial scale of spectral variations, and the need to obtain sufficent signal-to-noise to enable fitting of the broad component line profile.    

Filaments along the northern rim and parts of the southern rim of N103B exhibit a much simpler spectral behavior than the interior:  only the broad wings characteristic of Balmer-dominated shocks are observed, with minimal forbidden line emission.  The detection of these shocks offers a rare opportunity to measure the shock speed, degree of electron-ion equilibration and kinematic parameters of the forward shock in N103B.  Therefore, this paper is focused on characterizing the emission from these shocks, located at positions P1, P2, P3, and P4 in Figure~\ref{fig:fig3}.   

The spectra from positions M1, M2, and M3 mark locations of shocks exhibiting varying degrees of broad Balmer line emission mixed with varying strengths of forbidden line emission.  They are shown in Figure~\ref{fig:fig4} merely to demonstrate the range of emission conditions seen in the interior.  Note that the shocks at position M1 overlap the region where Li \etal\, (2017) detected redshifted broad Balmer line emission.  A detailed characterization of the radiative and partially radiative shocks in N103B is beyond the scope of this paper; we defer the in-depth discussion of these features (as well as possible abundance peculiarities if they arise from circumstellar wind material) to another paper.  We experimented with different extraction box locations, moving them until the sampled Balmer-dominated spectra had minimal contamination from nearby radiative shocks.

\section{Characteristics of the Broad Balmer Line Emission}

\begin{figure}[ht] 
 \centering
\includegraphics[width=7in]{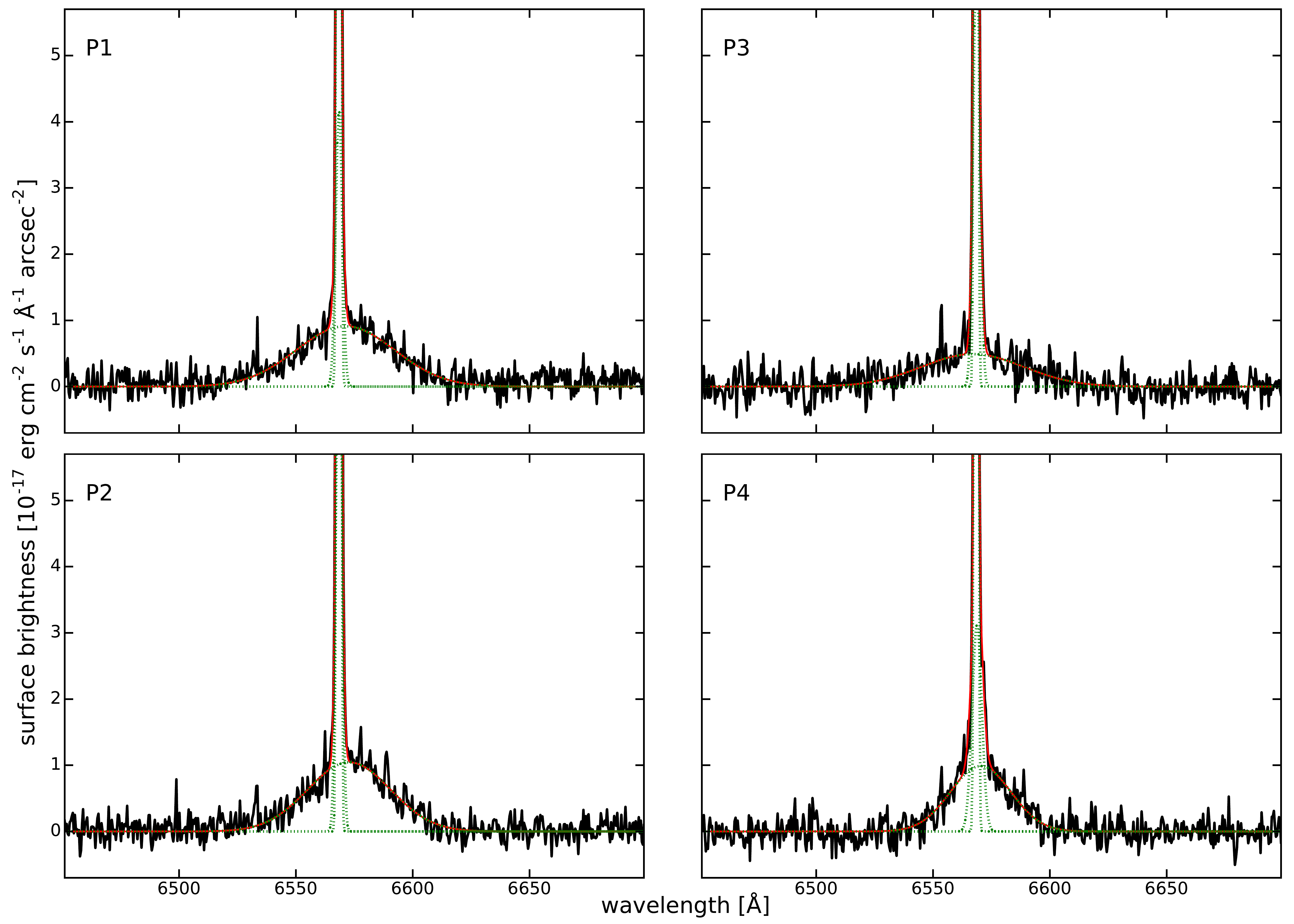}  
   \caption{ Background-subtracted \wifes\, spectra showing the H$\alpha$ profiles of Balmer-dominated shocks in N103B.  Each spectrum is fit with a 3-component Gaussian, with each component indicated by a green dotted line and the final combined fit indicated by the red solid line.  In each fit, the dotted line corresponding to the broad component fit overlaps the curve from the combined profile fit (red line).  All three emission line components are resolved (results from the fits are shown in Table~1).  The localized spikes seen in the blue wing of the broad component in spectra P1 -- P3 is an artifact of the sky subtraction.  A small amount of residual [N\,\textsc{ii}] $\lambda$6583 emission, not completely removed by the background subtraction, is seen in the spectrum of P2.}
   \label{fig:fig5}
\end{figure}

The \wifes\, spectra show that non-radiative (X-ray emitting) shocks are encountering partially neutral gas along the northern rim of N103B, giving rise to broad Balmer line emission at the top and bottom of the `3' (Figure~\ref{fig:fig5}).  The lack of forbidden line emission in the spectra of P1, P2, and P3 indicates that postshock cooling is negligible.  However, as mentioned earlier, a trace amount of [O\,\textsc{i}] $\lambda$6300 \AA\, emission remains in the spectrum of P3 even after all residual sky emission is subtracted out.    Balmer-dominated filamentary emission is also seen in the \wifes\, data cube running along the eastern rim, just outside the radiatively shocked interior clumps.  However, much of this emission is heavily smeared together with radiative and partially radiative shock emission from the interior, even at positions where Balmer-dominated emission is clearly not spatially coincident with radiative shock emission in higher resolution images (e.g., the HST image of N103B in the right panel of Figure~\ref{fig:fig3}).    The radiative shock emission shows multiple velocity components in [N\,\textsc{ii}], [S\,\textsc{ii}], [O\,\textsc{i}] and [O\,\textsc{iii}] (e.g., the spectra in M2 and M3), detailed interpretation of which would require disentangling radiative shock contribution to the broad Balmer lines.  Along the southern rim, a Balmer-dominated filament is seen at position P4 (Figures \ref{fig:fig2} and \ref{fig:fig3}), located between two brighter knots of radiatively shocked material.   

\subsection{H$\alpha$ Profiles and Line Fits} 

\begin{figure}[ht] 
 \centering
\includegraphics[width=7in]{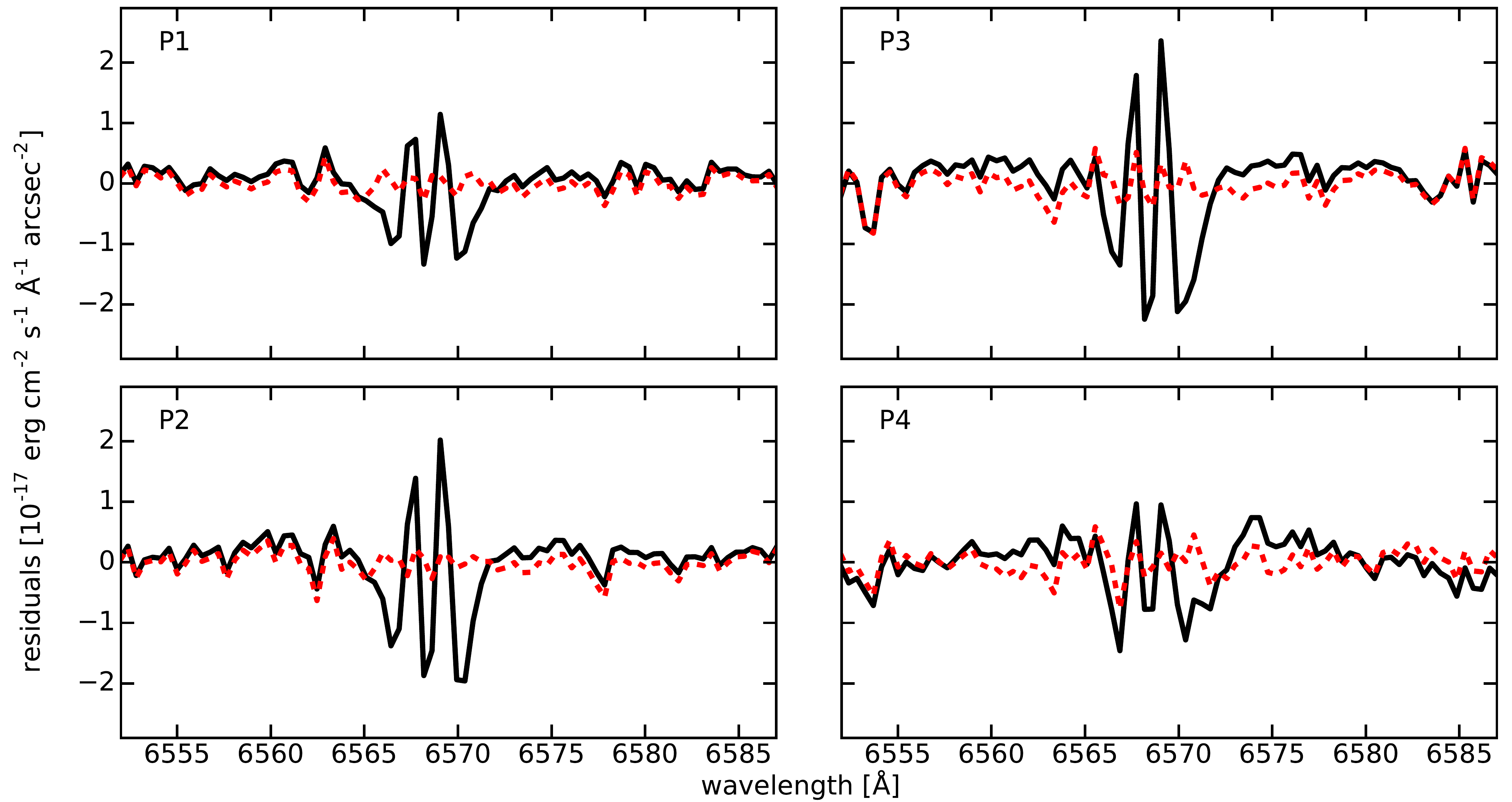}  
   \caption{ The residual emission resulting from subtraction of the two-Gaussian fit (solid black line) and three-Gaussian fit (dotted red line) from the H$\alpha$ line profiles.  The three-component fit produces markedly lower residuals near the centers of the profiles, demonstrating the presence of a third (intermediate width) component.  }
   \label{fig:fig6}
\end{figure}

At the beginning of our analysis of the H$\alpha$ line profiles we took the customary approach of fitting the profiles with two Gaussians.   We allowed the line width, centroid and intensity to vary for both the broad and narrow components (6 free parameters).    However, it soon became evident that this two-component model was inadequate for fitting the line profiles: the top of the broad component and the base of the narrow were systematically underfit.  We introduced a third Gaussian component (for a total of 9 free parameters) which dramatically improved the fit and eliminated the underfit H$\alpha$ flux.  All four positions, P1 -- P4, required a third Gaussian component to achieve satisfactory agreement with the data (see Figure~\ref{fig:fig5}).  Remarkably, all three line profile components have a larger width than the instrumental resolution of the red channel (43 \kms),  indicating that all three components are resolved.   The necessity of a third, intermediate component is further demonstrated in Figure~\ref{fig:fig6}, where we compare the residuals after subtraction of both a two-component Gaussian and three-component Gaussian fit from the data.  The residuals are substantially higher in the wings of the narrow H$\alpha$ line, indicating the presence of the intermediate component.  The reduced chisquared values of the 3-component Gaussian fits are (P1,P2,P3,P4) = (0.39, 0.38, 0.53 and 0.45), respectively, which are substantially lower than the 2-component Gaussian fits (P1,P2,P3,P4) = (0.58, 0.77, 1.08, 0.72) (note that the \wifes\, data reduction pipeline produces error bars that are slightly overestimated, resulting in reduced chisquareds lower than 1).     

In their calculation of the emission from narrow, intermediate and broad component neutrals in a 2000 \kms\, Balmer-dominated shock, Morlino \etal\, (2012) concluded that a spectral resolution of at least 70 \kms\, would be needed to successfully separate the three components.  By this measure, with an instrumental resolution of 43 \kms\, at H$\alpha$ we are given further assurance of the adequacy of \wifes\, in distinguishing the three components from one another.   

The results from our profile fits (corrected for the 43 \kms\, instrumental resolution) are shown in Table~1, with 1-$\sigma$ errors quoted.   The broad component H$\alpha$ widths of P1 (2260$\pm$105 \kms) and P3 (2360$\pm$190 \kms) are consistent with a single value to within the errors, while a slightly lower width (1900$\pm$80 \kms) is obtained for P2.  The average of these broad component widths, weighted by their uncertainties, is 2070$\pm$60 \kms.   The filament running through these three positions appears continuous in the higher resolution image (Figure~\ref{fig:fig3}, right), but bends sharply between P1 and P3.  This may reflect variation in preshock density along the northern rim, and corresponding variation in shock speed at that location.  

The widths of the narrow component H$\alpha$ lines are all near 45 \kms\, (Table~1) after correcting for the instrumental resolution.  These widths significantly exceed the nominal value of 10 \kms\, expected for warm gas in the interstellar medium.  This is consistent with similar enlargement observed in narrow component H$\alpha$ profiles of all other Balmer-dominated SNRs, both in the Milky Way (Sollerman \etal\, 2002) and LMC (Smith \etal\, 1994).  The enlarged narrow component widths observed these SNRs are now believed to reflect the preshock heating caused by dissipation of Alfv{\'e}n waves in a cosmic ray precursor (Morlino \etal\, 2013b).   

The H$\alpha$ broad components show a noticeable redshift in centroid along the northern edge of N103B.  The broad component is increasingly redshifted going from P1 (+130 \kms) to P2 (+200 \kms), with positions P3 and P4 showing only small velocity shifts.  A well-known result from the theory of Balmer-dominated shocks is that the magnitude and sign of the Doppler shift of the broad-component centroid is proportional to the shock viewing angle (Kirshner, Winkler \& Chevalier 1987; Ghavamian 2000; van Adelsberg\etal  2008).   This results from the net bulk velocity of the fast postshock neutrals, which reflects that of the shocked proton distribution behind the shock.  The redshifted P1 and P2 broad components indicate that these shocks are tilted slightly away from the line of sight.   

The spectrum from position M1 (top panel of Figure~\ref{fig:fig4}) shows the spectrum from a nearly face-on location (marked on Figure~\ref{fig:fig3}) located interior to filaments P1 -- P3.   Even after background subtraction, residual [O\,\textsc{i}] and [N\,\textsc{ii}] emission lines are clearly seen in the M1 spectrum.  These lines are broader than the corresponding lines in the background spectrum, indicating that they arise either from Balmer-dominated shocks transitioning into radiative shocks, or contamination from nearby radiative shock emission. The M1 spectrum shows a broad component having a width of approximately 1000 \kms, less than half that observed in P1 -- P3.  
    
The centroid of this broad component is redshifted to +400 \kms\, relative to the narrow component, consistent with the finding of Li \etal\, (2017) and suggesting that the Balmer-dominated shocks are being viewed at an angle $\sim$30$^\circ$ just inside the northern rim of N103B.  These shocks may form part of the same shell of emission which is observed tangent to the line of sight at positions P1 -- P3.  

\subsection{H$\beta$ Profile}

The broad H$\beta$ line was not detected in the blue channel data at positions P1, P2, and P3.  Subtracting the sky emission (using the same background regions as used for red channel spectra) produced false positive detections of broad H$\beta$ in the spectra of P1, P2, and P3.   This occurred due to the comparatively greater contamination of the blue channel spectra from background stars in the field of N103B.  The spectra of these stars contain broad absorption lines from the Balmer series, suggesting many are of early spectral type (A or F).  These absorption features were not eliminated by the stellar continuum removal described earlier for the red channel data.  This had the effect of embedding the H$\beta$ Balmer emission lines from P1, P2, and P3 in a local depressed background.  This depression then became a positive residual (i.e., a false broad emission feature) when the sky background was subtracted. 

\begin{figure}[ht] 
 \centering
\includegraphics[width=7in]{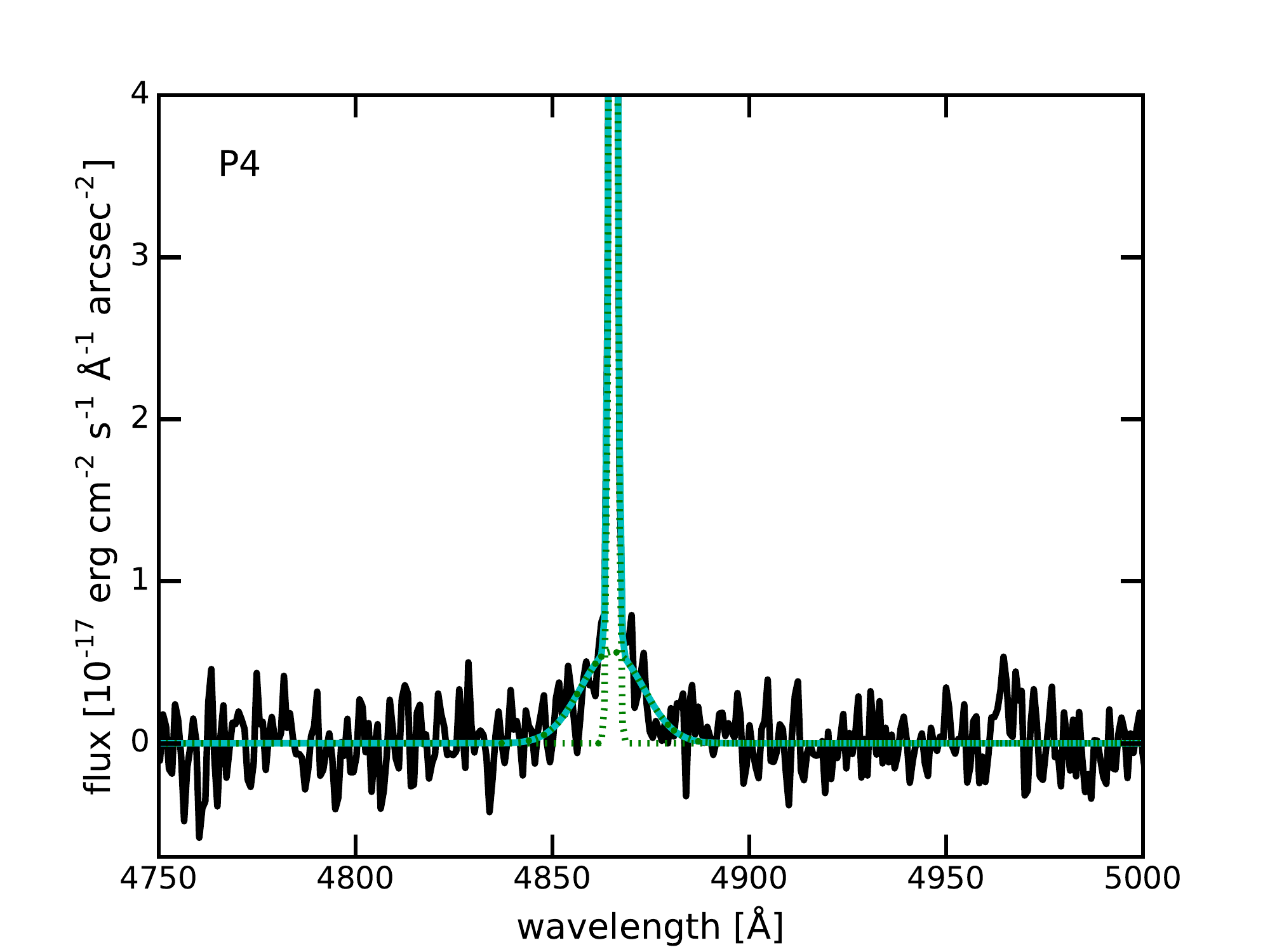}  
   \caption{ Fit to H$\beta$ profile for the Balmer-dominated shock at position P4.   The full fit is indicated by the cyan solid line, while the individual components (combined narrow and intermediate, and broad) are marked by the dotted green lines. }
   \label{fig:fig7}
\end{figure}

However, broad H$\beta$ is clearly detected from P4, even before sky subtraction.  The sky-subtracted spectrum is shown in Figure~\ref{fig:fig7}.  This detection is unsurprising, since the broad component width at that location is only half as large as that of the northern rim (P1 -- P3), and hence easier to detect.  At the spectral resolution of the blue channel (100 \kms\, at H$\beta$) the distinction between narrow and intermediate components is less clear, since the narrow component line profile is resolved in H$\alpha$ but not in H$\beta$, while the intermediate component is resolved in both H$\alpha$ and H$\beta$. There is insufficient S/N to fit the H$\beta$ profile with three components.  Therefore, we simply fit the profile with one quasi-narrow and one broad component, with both allowed to be free, with the caveat that the quasi-narrow component for H$\beta$ really represents a blend of the real narrow and intermediate components.  The resulting H$\beta$ broad component width is consistent with that of H$\alpha$ to within the errors (Table~1), giving confidence that it was reliably detected and properly fit.  The H$\beta$ broad-to-narrow ratio, 0.63$\pm$0.15, is nominally larger than that of H$\alpha$ (0.32$\pm$0.06 for the narrow+intermediate blend), though technically similar to within the errors.  A larger ratio is expected for H$\beta$ due to the lower effectiveness of Ly $\gamma$ trapping on enhancing the narrow H$\beta$ compared to Ly $\beta$ trapping enhancing the narrow H$\alpha$ (Chevalier, Kirshner \& Raymond 1980; Ghavamian\etal 2001, 2002).  

\subsection{The Balmer Decrement}

Since broad H$\beta$ emission was not detected from the northern rim of N103B, we fit the narrow line with a single Gaussian to estimate its flux and width at that location (Table~1).   After correction for instrumental broadening (100 \kms), the weighted average widths of the H$\beta$ single profiles is 52 \kms\, for all four positions.  This is slightly larger than those of the corresponding narrow-only H$\alpha$ lines (45 \kms).  This is consistent with a small contribution from an unresolved intermediate velocity component to the H$\beta$ line profiles, further evidence that the intermediate velocity component exists.  On the other hand, since broad H$\beta$ was detected from the slower shocks at P4, we were able to estimate the broad Balmer decrement at that location separately from that of the other (blended) narrow and intermediate components.  First we added the fit H$\alpha$ fluxes from the narrow and intermediate components, then divided the result by the fitted flux from the single narrow H$\beta$ line.   The resulting Balmer decrement (narrow+medium) is $I(H\alpha)_{n+m}/I(H\beta)_{n+m}$ = 5.14$\pm$0.27, while the Balmer decrement for the broad component alone is $I(H\alpha)_{b}/I(H\beta)_{b}$ = 2.67$\pm$0.67.  Note that the uncertainty of the broad Balmer decrement does not include the systematic uncertainty in the baseline level, which affects the measured flux in the broad H$\beta$ line.  

The observed Balmer decrements need to be corrected for the interstellar extinction toward N103B.   The relationship between the observed Balmer decrement and the intrinsic Balmer decrement can be expressed in terms of the visual extinction parameter $A$ via
\begin{equation}
\left[\frac{I(H\alpha)}{I(H\beta)}\right]_{0}\,=\,\left[\frac{I(H\alpha)}{I(H\beta)}\right]_{obs} 10^{ \,-0.4 \,(A_{H\beta}\,-\,A_{H\alpha})      }
\end{equation} 
We estimated $A_{H\beta}$ and $A_{H\alpha}$ from the average LMC extinction curves of Weingartner \& Draine (2001) with a carbon abundance
$b_c$\,=\,2$\times$10$^{-5}$.  This gives $A_{H\beta}\,\,\approx\,$1.44$\times$10$^{-22}\,N_H$ mag and $A_{H\alpha}\,\,\approx\,$8.40$\times$10$^{-23}\,N_H$ mag, where $N_H$ is the H column density along the line of sight.   
Lewis\etal  (2003) modeled $N_H$ from their \chan\, spectra of N103B, and found $N_H\,\approx\,$3.5$\times$10$^{21}$ cm$^{-2}$.   This gives $\left[\frac{I(H\alpha)}{I(H\beta)}\right]_{0}\,\approx\,0.824\,\left[\frac{I(H\alpha)}{I(H\beta)}\right]_{obs}$, with reddening corrected Balmer decrements of $\left[I(H\alpha)_{n+m}/I(H\beta)_{n+m}\right]_0$ = 4.23$\pm$0.22 and $\left[I(H\alpha)_{b}/I(H\beta)_{b}\right]_0$ = 2.20$\pm$0.55.   The latter Balmer decrement is lower than the minimum value of 3.0 predicted for a gas optically thin in the Lyman lines.  This may be due to a slight overestimate in the H$\beta$ broad-component flux in our line fits.

The Balmer decrements for positions P1, P2, and P3 can only be estimated for the combined narrow and intermediate lines, given the blending of these lines in H$\beta$ and the lack of detected broad H$\beta$ 
at these positions.    The observed Balmer decrements for the combined narrow and intermediate lines are (P1,P2,P3) = (3.9$\pm$0.4, 3.8$\pm$0.3 and 2.6$\pm$0.2), respectively.   Corrected for interstellar extinction, these are (P1,P2,P3) = (3.5$\pm$0.3, 3.4$\pm$0.3, 2.3$\pm$0.2).   

\section{Interpretation of the Balmer Line Profiles: Broad-to-Narrow Ratios and Balmer Decrement}

The broad-to-narrow ratios reported in Table~1 are presented for the cases where the narrow and intermediate components are blended (in which case the sum of the narrow and intermediate fluxes is used), and the other where the two components are resolved (in which case only the narrow component flux is used).  The former broad-to-narrow ratios are 2-3 times smaller than values found in Balmer-dominated remnants of similar broad-component width, such as Tycho's SNR, SN 1006, and Kepler's SNR in our galaxy and SNR 0519$-$69.0 in the LMC (see van Adelsberg\etal  2008 and Ghavamian\etal 2013 and references therein for summaries on the measured broad-to-narrow ratios in the literature).   Even using solely the narrow component flux (eighth column in Table~1), the broad-to-narrow ratios increase $\sim$15\%, well short of what is needed to bridge the difference between the N103B ratios and the other Balmer-dominated SNRs.   The cause of the low broad-to-narrow ratios has still not been conclusively determined, but one possible cause is collisional excitation of narrow H$\alpha$ within a spatially unresolved cosmic ray precursor (Raymond \etal\, 2011, Morlino et al. 2012).

Modeling the low broad-to-narrow ratios in N103B is problematic because they are smaller than predicted by any numerical model, such as those of Ghavamian\etal (2001, 2002, 2013) and van Adelsberg\etal  (2008).  The discrepancy most likely occurs because the numerical models do not include emission from neutrals ahead of the shock (Raymond\etal 2011; Morlino\etal 2012).   Heating in the cosmic ray precursor can collisionally excite narrow-component Balmer line emission from slow neutrals (Raymond\etal 2011), while excitation of hydrogen in the neutral return flux can excite intermediate component Balmer emission ahead of the shock. These effects may play a more significant role in N103B than for previously modeled remnants such as Tycho's SNR and SN 1006.  Therefore, the method  of simultaneously determining the downstream electron-proton equilibration and shock speed cannot be applied here as it was in Tycho's SNR and SN 1006 (Ghavamian\etal 2001, 2002).  

In the absence of reliable models for the broad-to-narrow ratios in N103B, we can at least use model predictions for broad-component FWHM as a function of shock speed to bracket the range of shock speeds.  The latest models are those of Morlino\etal  (2013), who presented an empirically fitted relation between the broad-component FWHM and shock speed for different values of the initial downstream equilibration, $\beta_{down}$.  This relation was derived assuming negligible loss of shock energy to cosmic ray acceleration.  Inserting our measured broad-component widths into these relations, we obtained the range of shock speeds between $\beta_{down}\,=\,$0.01 and $\beta_{down}\,=\,$1.0 (Table~2). Furthermore, Morlino\etal (2013) showed that in cases where the ratio of cosmic ray pressure to ram pressure ($\epsilon_{CR}\,\equiv\,P_{CR}/\rho_{0,ion}\,V_{sh}^2$) is less than 20\%, the broad FWHM for a given shock speed is a linear function of $\epsilon_{CR}$.  Using the Morlino\etal (2013) results, the change in broad-component width, $\Delta V_{FWHM}(\epsilon_{CR})$, in going from a shock with negligible cosmic ray acceleration efficiency and having broad-component width $V_{FWHM,0}$ to a shock having non-zero acceleration efficiency $\epsilon_{CR}$, is
\begin{equation}
\Delta\,V_{FWHM}(\epsilon_{CR})\,\approx\,\frac{2000}{V_{FWHM,0}}\,\epsilon_{CR}
\end{equation}
where $V_{FWHM}$ is measured in \kms. If we conservatively assume $\epsilon_{CR}\,\sim\,$0.2 for the N103B shocks, then if the downstream electron-ion equilibration is low ($\beta\,\sim\,$0.01, a value close to those inferred for nearly all Balmer-dominated shocks exceeding 1000 \kms), we can expect the broad-component widths to have been reduced by at most 17\% -- 20\% from the values they would have for negligible cosmic ray acceleration efficiency.   Proper motion and spectroscopic studies of Balmer-dominated shocks have thus far shown that a small amount of cosmic ray acceleration is required to account for the enlarged widths of the narrow component H$\alpha$ lines (and possibly enhanced emission in the narrow component, as mentioned above).  However, there is broad consistency between X-ray and optical proper motions and the shock speeds obtained from Balmer-dominated shock models excluding cosmic ray acceleration (e.g., Winkler, Gupta \& Long 2003 for SN 1006, Hovey\etal  2015 for SNR 0509-67.5, and Williams\etal  2016 for Tycho's SNR).  The lack of significant non thermal emission in the X-rays and weakness of the radio emission (Mathewson\etal\ 1983) in the four known Balmer-dominated LMC remnants and in N103B is further evidence for the lack of efficient cosmic ray acceleration (i.e., $\epsilon_{CR}\,\gtrsim\,$0.5).  Therefore, the shock velocity ranges reported in Table~2 are unlikely to be underestimated by more than 20\%. 

The most recent theoretical predictions of the Balmer decrements of non-radiative shocks exceeding 1000 \kms\, (an appropriate regime for  N103B) have been made by Tseliakhovich \etal\, (2012).  At these shock speeds, excitation and ionization of hydrogen is dominated not by electrons but by protons ($E\,\gtrsim\,$5 keV).  These authors computed the Balmer decrement under the extreme assumptions of Case A (gas optically thin to Lyman line radiation) and Case B (gas optically thick to Lyman line radiation).  The former is appropriate for computing the broad component Balmer decrement, since the large width of the line profile results in negligible absorption of Lyman line photons (Chevalier, Kirshner \& Raymond 1980; Ghavamian \etal\, 2001), and hence little enhancement of the Balmer decrement. The latter case is appropriate for computation of the narrow  component (and to a lesser degree, the intermediate component) Balmer decrement.       
Comparing our observed Balmer decrement for P4 with the predictions of Tseliakhovich \etal\, (2012, Figure~9 of that paper), we unfortunately find that the proton temperature predicted for the P4 shocks (2.6 keV to 3.3 keV between the limits of minimal and full equilibration, respectively; Table~2) fall mostly out of range of their theoretical calculations, which are all computed from 5 keV onward.  However, their models show a steep monotonic rise in Balmer decrement at low proton impact energies, rising from less than 3.0 near 10 keV to values of 4.0 and 4.9 at 5 keV for Case A and Case B, respectively.  Although the precise values below 5 keV are not calculated, these theoretical Balmer decrements are consistent with the value of 4.2$\pm$0.2 we measure for P4.  In addition, the minimum Case A Balmer decrement calculated by Tseliakhovich \etal\, (2012) is 2.6, which lies at the upper bound of our measured value for the broad component of P4, 2.2$\pm$0.6, though the minimum is reached for 10 keV in their models rather than 3 keV as appropriate to our case.   The Balmer decrements of the narrow+intermediate components for the P1, P2 and P3 shocks are consistent with those calculated by Tseliakhovich \etal\, (2012) for proton impact energies of 2000 \kms\, shocks ($\sim$8 keV) , though better constraints should be possible with higher signal-to-noise data.    

\section{Interpretation of the Balmer Line Profiles: Intermediate Component}

One of the most exciting results from our WiFeS observations of N103B is the detection of the intermediate width H$\alpha$ component.   Such a component was detected in high resolution
spectra of Balmer-dominated shocks in Tycho's SNR (Ghavamian\etal  2000), and possibly in high resolution spectra of SNR 0509$-$67.5 (Smith\etal  1994).  The line width of the narrow and intermediate components
found by Ghavamian\etal  (2000) in Knot g of Tycho's SNR were 44 \kms\, and 150 \kms, respectively, which is very similar to that of  shocks for positions P1 -- P3 in N103B (Table~1).  The broad-component width 
measured for Knot g ($\sim$1800 -- 2100 \kms; Ghavamian 1999; Ghavamian\etal  2001) is also similar to those we observed in positions P1-P3 in N103B.   One major exception is position P4, where the width of the broad component is only half that of the other three positions, but where the width of the intermediate component is nearly twice as large as the other three positions.   

The presence of an intermediate neutral population was predicted theoretically by Morlino\etal  (2012), who found that the fast neutrals crossing back upstream should deposit energy ahead of the shock in a fast neutral precursor.  One observational consequence is the production of Balmer line emission having a width of a few hundred \kms\, (i.e., wider than the narrow component but narrower than the broad component).   Morlino\etal  (2012) found that the relative importance of the narrow, intermediate, and broad components depends on the shock velocity, preshock neutral fraction, and on the electron-ion temperature equilibration level upstream.  The P1 and P3 shocks  have very similar line widths to within the errors, so if we assume low postshock equilibration ($\beta\,\lesssim$0.1), their speeds are approximately 2500 \kms.  From the Morlino\etal (2012) models for this shock speed and equilibration, the intermediate component line width is predicted to lie between 290 \kms\, and 320 \kms\, for $m_e/m_p\,\leq\,\beta\,\leq\,$0.1. Although there is lack of close agreement between the model predictions and the observed intermediate component width of 145 \kms, there is a rough order of magnitude consistency between the two.     

Curiously, the intermediate component model assuming a fully equilibrated plasma ($\beta\,=\,$1 both upstream and downstream for $V_{sh}\,\approx\,$3400 \kms) matches the width observed for the intermediate component.  However, fully equilibrated shocks exceeding 1000 \kms\, have yet to be observed in SNRs, and the trend of declining postshock equilibration with shock speed (Ghavamian\etal  2007; van Adelsberg \etal\, 2008; Ghavamian \etal\, 2013)  has been consistently found to  describe shocks over a wide range of supernova remnants, so the full equilibration scenario seems very unlikely for the P1 -- P3 shocks.  Finally, the models of Morlino\etal  (2012) come much closer to matching the intermediate component widths for P4 (218 \kms) than the other three positions, with predictions falling between 170 \kms\, to 210 \kms\, as long as minimal and maximal equilibration are ruled out.  This is not unreasonable, though better agreement should still be possible with improved shock models.  

Morlino\etal  (2012) pointed out that the large width of the intermediate component observed in Knot~g by Ghavamian\etal  (2000) may be caused by geometric broadening induced by the
complex structure of Knot~g.    However, the P1 -- P3 shocks in N103B are interacting with a seemingly less complicated environment than Tycho, so the strong similarity we measure in the narrow, intermediate
and broad component widths of the two SNRs is striking, and may weaken the appeal of the geometric broadening argument.  The Morlino\etal  (2012) models of the intermediate component do not include 
effects of the cosmic ray precursor, which is expected to play a significant role in the heating and density structure of the neutral gas.  It is possible that the agreement between the observations and models will improve once the back reaction from the cosmic rays is included.

\section{Implications for the Evolution of N103B}

The shock velocities derived for N103B can be used to estimate the age of the SNR. Excluding the shock velocity along the southern rim (P4) where the remnant has undergone greater deceleration than the north, and taking the average of the shock speeds for P1, P2, and P3 weighted by their uncertainties, then the average shock speed of the least decelerated Balmer-dominated shocks is $V_{sh}\,\approx\,$2070$\pm$60 \kms, assuming $\beta\,\approx\,$0.01.  If $\beta$ is closer to 0.1, a higher shock speed results, but it would only be $\sim$5\% higher than the quoted value.   The SNR appears approximately circular in the archival \chan\, image, with radius approximately 15${\arcsec}$ (Lewis \etal\, 2003) (this also approximately matches the radius measured along the north-south direction from the HST images, Li\etal  (2017)).   For a distance of 50 kpc  this translates to a radius of 3.6 pc.    For a SNR in the Sedov stage of evolution, the age is $\tau\,=\,2R/5V_{sh}$, which for the quoted parameters gives $\tau\,\approx\,$685$\pm$20 years for N103B.    An upper limit on the age of N103B can be derived under the assumption of free expansion throughout the lifetime of the SNR, which gives $\tau\,=\,R/V_{sh}$\,=\,1710$\pm$50 years.   Rest \etal\, (2005) found a light echo age of 860 years under the assumption that the scattering dust sheet lies in the plane of the sky, with a lower limit of 380 years if the
dust sheet lies at an angle up to  60$^{\circ}$ from the plane of the sky.  Comparing these values to our estimates, a high inclination dust sheet geometry is clearly ruled out, as well as the free expansion age of 1710 years.  Considering the uncertainties (e.g., variations in forward shock radius for N103B, dust sheet inclination), the Sedov age of 685 years is consistent with the age of 860 years quoted by Rest \etal\, (2005).  This is also consistent with the results of Williams \etal\, (2014), who calculated the mass of shocked dust in N103B and concluded that a few solar masses of material had been swept up by the SNR.  This indicates that N103B is past the free expansion phase ($M_{swept}\,<\,M_{ej}$), and beginning to enter the Sedov-Taylor phase ($M_{swept}\,\gg\,M_{ej}$).    

\section{Discussion}
Recent numerical models have made enormous strides in addressing the missing physics of Balmer-dominated shocks, thorough kinetic treatment of ion-neutral momentum exchange, inclusion of back pressure from cosmic rays, inclusion of heating from the neutral return flux, etc. However, since the work of Ghavamian\etal (2001),
we know of no attempt to model the detailed conversion of Ly $\beta$ photons into H$\alpha$ in Balmer-dominated shocks.   The recent models of Tseliakhovich \etal (2012), while including proton impact excitation, do not include the radiative transfer conversion.  It is usually assumed that Case B conditions hold in the narrow component optically thick conditions in Lyman lines and Case A in the broad (optically thin conditions in the Lyman lines) (Heng\etal  2007; van Adelsberg\etal  2008; Morlino\etal  2012, 2013).  However, this fails to capture the sensitivity of the H$\alpha$ broad-to-narrow ratio to the preshock neutral fraction of hydrogen, which was modeled by Ghavamian\etal  (2001, 2002).   

The sensitivity of the broad-to-narrow ratio to preshock neutral fraction and cosmic ray acceleration efficiency is yet to be studied theoretically in full detail. 
The Balmer models show that a moderate efficiency of cosmic ray acceleration, $\epsilon_{CR}\,\sim$0.2, is necessary for widening the narrow component line profile and collisionally exciting H$\alpha$ in the cosmic ray precursor (Vink \etal\, 2010; Raymond \etal\, 2011).   The broad component width and broad-to-narrow flux ratio, formerly a diagnostic of shock speed, neutral fraction and electron-ion equilibration, have now been shown to depend on $\epsilon_{CR}$ as well.  There is a need for additional constraints, such as estimates of the shock speed from proper motion measurements (Morlino\etal  2013).  

 \section{Summary}

We have presented results from integral field spectroscopy of Balmer-dominated shocks in the supernova remnant N103B in the LMC with the WiFeS instrument.  Our main results are as follows:

1. We have detected broad H$\alpha$ and H$\beta$ emission characteristic of Balmer-dominated shocks along the rim of N103B.  The broad component widths range from values as high as 2350 \kms\, FWHM along the northern edge down to as low as 1300 \kms\, along the southern edge.  These widths correspond to shock speeds around 2100 \kms\, and 1200 \kms, respectively, assuming low electron-ion equilibration at the shock front.  

2. We find that the H$\alpha$ line profiles are not adequately described by the usual two gaussian fit, but rather require a three component fit consisting of narrow, intermediate and broad lines.  The intermediate component H$\alpha$ lines exhibit widths ranging from around 125 \kms\, to 220 \kms\, FWHM, significantly wider than the narrow component (around 45 \kms) and much narrower than the broad component (around 2000 \kms).   The detection of an intermediate component is  consistent with the predictions of the latest models of Balmer-dominated shocks (Morlino \etal\, (2012, 2013a, 2013b)), which predict that a third population of neutrals having a temperature intermediate between the narrow and broad components is formed ahead of the shock when broad component neutrals leak upstream and undergo charge exchange with preshock ions.   There is an order of magnitude agreement between the observed and predicted intermediate component widths, though more improved shock modeling may improve the agreement in the future.


3. The forward shock velocity of N103B, averaged over several locations around the rim and weighted by the uncertainty in broad component width, is 2070$\pm$60 \kms\, assuming low electron-ion equilibration ($\sim$1\%). This gives an SNR age of around 685 years under the assumption that it is in the Sedov stage of evolution, consistent with the range 380$-$860 years estimated from the study of its light echoes.   

\paragraph{Acknowledgments.} This research has made use of the following \textsc{python} modules: \textsc{astropy}, a community-developed core \textsc{python} package for Astronomy (Astropy Collaboration et al. 2013), \textsc{aplpy}, an open-source plotting package for \textsc{python} hosted at \url{http://aplpy.github.com}, and \textsc{mpfit}, a \textsc{python} script that uses the Levenberg-Marquardt technique (Mor{\'e} 1978) to solve least-squares problems, based on an original \textsc{fortran} code part of the \textsc{minpack-1} package.

This research has also made use of the \textsc{aladin} interactive sky atlas (Bonnarel et al. 2000), of \textsc{saoimage ds9} (Joye \& Mandel 2003) developed by Smithsonian Astrophysical Observatory, of NASA’s Astrophysics Data System, and of \textsc{montage}, funded by the National Science Foundation under Grant Number ACI-1440620 and previously funded by the National Aeronautics and Space Administration’s Earth Science Technology Office, Computation Technologies Project, under Cooperative Agreement Number NCC5-626 between NASA and the California Institute of Technology. This work has also made use of data from the European Space Agency (ESA) mission {\it Gaia} (\url{https://www.cosmos.esa.int/gaia}), processed by the {\it Gaia} Data Processing and Analysis Consortium (DPAC, \url{https://www.cosmos.esa.int/web/gaia/dpac/consortium}). Funding for the DPAC has been provided by national institutions, in particular the institutions participating in the {\it Gaia} Multilateral Agreement.

The authors would like to thank the anonymous referee for helpful comments that helped improved the content and presentation of this paper.  The work of P. G. was provided by grants HST-GO-12545.08 and HST-GO-14359.011.  P. G. would also like to thank funding support from the Distinguished Visitors Program of the Australian National University, the ANU CAASTRO node, and the hospitality of the staff at Mt.\ Stromlo observatory, where most of this paper was written. IRS acknowledges support from the Australian Research Council Laureate Grants FL0992131 and FT160100028. PFW also acknowledges the hospitality of Mt.\ Stromlo Observatory staff during the period when planning for the \wifes\, observations was initiated, and the financial support of the NSF through grant AST-098566. Parts of this research were conducted by the Australian Research Council Centre of Excellence for All-sky Astrophysics (CAASTRO), through project number CE110001020.

\clearpage

\begin{deluxetable}{ccccccccc}
\tabletypesize{\footnotesize}
\tablewidth{0pt}
\tablecolumns{7}
\tablecaption{Balmer-Dominated Spectral Parameters for N103B}
\tablehead{
\colhead{Position} & \colhead{$V_{\rm FWHM}$(b) } & \colhead{$\Delta V$(b) } &
\colhead{$V_{\rm FWHM}$(m)\tablenotemark{a} }  &  \colhead{$\Delta V$(m)} &  \colhead{$V_{\rm FWHM}$(n)\tablenotemark{a}}	& 	\colhead{I$_b$/(I$_m$\,+\,I$_n$)}  &   \colhead{I$_b$/I$_n$}  &  \colhead{I$_{total}$}\\
\colhead{}	&	\colhead{(km/s)}	&	\colhead{(km/s)}	&	\colhead{(km/s)}	&	\colhead{(km/s)}	&	\colhead{(km/s)}		&	\colhead{}		&	\colhead{}		&	\colhead{(10$^{-15}$ ergs/cm$^2$/s/arcsec$^2$)  }  }
\startdata
\cutinhead{H$\alpha$ Balmer-Dominated Spectral Parameters\tablenotemark{b} (Red Channel)}
1	&	2260$\pm$105	&	129$\pm$42	&	146$\pm$15	&	3$\pm$3	&	48.7$\pm$0.5	&	0.41$\pm$0.11	&	0.47$\pm$0.03	&	1.64$\pm$0.05\\	
2	&	1900$\pm$80	&	200$\pm$34	&	124$\pm$8	&	4$\pm$1	&	45.9$\pm$0.5	&	0.29$\pm$0.06	&	0.35$\pm$0.02	&	2.05$\pm$0.06\\	
3	&	2360$\pm$190	&	$-$66$\pm$78		&	144$\pm$7	&	10$\pm$2	&	42.6$\pm$0.5	&	0.23$\pm$0.04	&	0.29$\pm$0.03	&	1.43$\pm$0.05\\
4	&	1280$\pm$100\tablenotemark{c}	&	58$\pm$34		&	218$\pm$17 		&	14$\pm$4	&	45.1$\pm$0.04	&	0.32$\pm$0.06	&	0.38$\pm$0.05	&	1.29$\pm$0.04\\
\cutinhead{H$\beta$ Balmer-Dominated Spectral Parameters (Blue Channel) }
1	&	\nodata	 &	\nodata 	&	\nodata	&	\nodata	&	57.0$\pm$1.0	&	\nodata\tablenotemark{c}	&	\nodata\tablenotemark{c} 	&	0.30$\pm$0.03 \\	
2	&	\nodata	 &	\nodata	&	\nodata	&	\nodata	&	59.4$\pm$0.7	&	\nodata	&	\nodata 	&	0.41$\pm$0.03 \\	
3	&	\nodata	&	\nodata	&	\nodata	&	\nodata	&	46.8$\pm$0.5	&	\nodata	&	\nodata  	&	0.45$\pm$0.03  \\	
4	&	1200$\pm$190\tablenotemark{a,d}	&	-15$\pm$72	&	\nodata	&	\nodata	&	62.6$\pm$2.1	&	0.63$\pm$0.15	&	\nodata	&	0.31$\pm$0.02\\	
\enddata
\tablenotetext{a}{Corrected for the 43 \kms\, instrumental broadening of the red channel or 100 \kms\, of the blue channel}  
\tablenotetext{b}{3-component gaussian fit consisting of narrow (n), intermediate (m) and broad (b) components.}
\tablenotetext{c}{No broad H$\beta$ detected in spectra of Positions 1, 2 and 3, so profiles are fit with a 1-component gaussian consisting of a single quasi-narrow (n) component.  }
\tablenotetext{d}{2-component gaussian fit consisting of a quasi-narrow and broad component.  The quasi-narrow component includes both the narrow and intermediate components, since they are blended at the 100 \kms\, instrumental resolution of the blue channel.  }

\end{deluxetable}

\begin{deluxetable}{ccc}
\tablewidth{300pt}
\tablecaption{Derived Shock Velocities (km/s) for Balmer-Dominated Shocks in N103B\tablenotemark{a}}
\tablehead{
\colhead{Position} & \colhead{V$_{sh}(\beta\,=\,0.01)$ } & \colhead{V$_{sh}(\beta\,=\,1)$ }  }
\startdata
1	&	2360$\pm$90	&	3330$\pm$125\\
2	&	1850$\pm$65	&	2630$\pm$90\\
3       &       2530$\pm$145     &      3560$\pm$205\\
4       &       1160$\pm$90    &       1670$\pm$130\\
\enddata
\tablenotetext{a}{Derived from the H$\alpha$ broad component width using the $V_{sh}$ vs $V_{FWHM}$ relation of Morlino \etal\, 2013b} 
\tablenotetext{b}{$\beta$ is the degree of downstream electron-ion equilibration, $T_e/T_p$.  }

\end{deluxetable}

\clearpage

\begin{references}  

\reference{B91} Blair, W. P., Long, K. S. \& Vancura, O. 1991, \apj, 366, 484
\reference{B12} Blasi, P., Morlino, G., Bandiera, R., Amato, E. \& Caprioli, D. 2012, \apj, 755, 121
\reference{B00} Bonnarel, F., \etal\, 2000, \aap, 143, 33
\reference{C80} Chevalier, R. A., Kirshner, R. P. \& Raymond, J. C. 1980, \apj\, 235, 186
\reference{CR78} Chevalier, R. A. \& Raymond, J. C. 1978, \apj\, 225, L27
\reference{C14} Childress, M. , Vogt, F., Nielsen, J. \& Sharp, R. 2014, {\it Astrophysics Source Code Library}, \apss, 349, 617
\reference{D07} Dopita, M. A., Hart, J., McGregor, P., Oates, P., Bloxham, G. \& Jones, D. 2007, \apss, 310, 255 
\reference{D10} Dopita, M. A. \etal\, 2010, \apss,  327, 245
\reference{G00} Ghavamian, P., Raymond, J., Hartigan, P. \& Blair, W. P. 2000, \apj, 535, 266  
\reference{G01} Ghavamian, P., Raymond, J., Smith, R. C. \& Hartigan, P. 2001, \apj, 547, 995  
\reference{G02} Ghavamian, P., Winkler, P. F., Raymond, J. C. \& Long, K. S. 2002, \apj, 572, 888
\reference{G07} Ghavamian, P.,  Laming, J. M. \& Rakowski, C. E. 2007, \apj, 654, L69
 
\reference{H07} Heng, K., van Adelsberg, M., McCray, R. \& Raymond, J. C. 2007, \apj, 668, 275
\reference{H10} Heng, K. 2010, Publ. Astr. Soc. Australia, 27, 32
\reference{H15} Hovey, L, Hughes, J. P. \& Eriksen, K. 2015, \apj, 809, 119
\reference{H95} Hughes, J. P. \etal\, 1995, \apjl, 444, L71
\reference{Iben1984} Iben, Jr., I. \& Tutukov, A. V. 1984, \apjs, 54, 335
\reference{JM03} Joye, W. A. \& Mandel, E. 2003, in Astronomical Data Analysis Software and Systems XII ASP Conference Series, Vol. 295, H. E. Payne, R. I. Jedrzejewski, and R. N. Hook, eds., p.489
\reference{L03} Lewis, K.  T., Burrows, D. N., Hughes, J. P., Slane, P. O., Gamire, G. P. \& Nousek, J. A. 2003, \apj, 582, 770
\reference{L17} Li, C.-J., \etal  2017, \apj, 835, 85
\reference{L11} Lopez, L. A., Krumholz, M. R., Bolatto, A. D., Prochaska, J. X. \& Ramirez-Ruiz, E. 2011, \apj, 731, 91
\reference{M09} Markwardt, C. B. 2009, "Non-Linear Least Squares Fitting in IDL with MPFIT," in proc. 
Astronomical Data Analysis Software and Systems XVIII, Quebec, Canada, ASP Conference Series, Vol. 411, eds. D. Bohlender, P. Dowler 
\& D. Durand (Astronomical Society of the Pacific: San Francisco), p. 251-254
\reference{M83} Mathewson, D. S., Ford, V. L., Dopita, M. A., Tuohy, I. R., Long, K.S. \& Helfand, D. J. 1983, \apjs, 51, 345
\reference{M78} Mor{\'e}, J. 1978, "The Levenberg-Marquardt Algorithm: Implementation and Theory," in Numerical Analysis, vol. 630, ed. G. A. Watson (Springer-Verlag: Berlin), p. 105
\reference{M12} Morlino, G., Blasi, P., Bandiera, R. \& Amato, E. 2012, \apj, 760, 137
\reference{M13a} Morlino, G., Blasi, P., Bandiera, R. \& Amato, E. 2013, \aap, 558, A25 (Morlino \etal  2013a)
\reference{M13b} Morlino, G., Blasi, P., Bandiera, R., Amato, E. \& Caprioli, D. 2013, \apj, 768, 148 (Morlino \etal  2013b)
\reference{R11} Raymond, J. C., Vink, J., Helder, E. A. \& de Laat, A. 2011, \apj, 731, L14
\reference{R05} Rest, A., Suntzeff, N.B., Olsen, K. et al.\ 2005, Nature, 438, 1132
\reference{S16} Sankrit, R., Blair, W. P., Long, K. S., Williams, B. J., Borkowski, K. J., Patnaude, D. J., \& Reynolds, S. P. 2016, \apj, 817, 36
\reference{S94} Smith, R. C., Raymond, J. C., \& Laming, J. M. 1994, \apj\, 420, 286
\reference{S02} Sollerman, J., Ghavamian, P., Lundqvist, P. \& Smith, R. C. 2003, \aap, 407, 249
\reference{T12} Tseliakhovich, D., Hirata, C. M. \& Heng, K. 2012, \mnras, 422, 2357
\reference{T82} Tuohy, I. R., \etal  1982, \apj, 261, 473
\reference{V08} van Adelsberg, M., Heng, K., McCray, R, \& Raymond, J. C. 2008, \apj, 689, 1089
\reference{V10} Vink, J., Yamazaki, R., Helder, E. A. \& Schure, K. M. 2010, \apj, 722, 1727
\reference{Webbink1984 }Webbink, R.~F. 1984, \apj, 277, 355
\reference{W03} Weingartner, J. C., \& Draine, B. T. 2001, \apj, 548, 296
\reference{Whelan1973} Whelan, J. \& Iben, I. J. 1973, ApJ, 186, 1007
\reference{Williams, B. J., \etal\, 2016, \apj, 823, 32 }
\reference{WGL03} Winkler,  P. F., Gupta, G. \& Long, K. S. 2003, \apj, 585, 324
\reference{Y99} Yamaguchi, S., Koyama, K., Tomida, H., Yokogawa, J. \& Tamura, K. 1999, \pasj, 51, 13

\end{references}
\end{document}